\documentclass[%
 reprint,
 amsmath,amssymb,
 aps,
 pra,
]{revtex4-2}

\usepackage{xcolor}
\usepackage{braket}
\usepackage{graphicx}
\usepackage{dcolumn}
\usepackage{bm}

\begin{document}

\preprint{APS/123-QED}

\title{Polarization diverse true heterodyne receiver for continuous variable quantum key distribution}

\author{Daniel Pereira}
 \email{danielfpereira@ua.pt}
 \affiliation{ Instituto de Telecomunicações, University of Aveiro, Campus Universitário de Santiago, 3810-193, Aveiro, Portugal}%
\affiliation{Department of Electronics, Telecommunications and Informatics, University of Aveiro, Campus Universitário de Santiago, 3810-193, Aveiro, Portugal}%

\author{Nuno A. Silva}
\affiliation{ Instituto de Telecomunicações, University of Aveiro, Campus Universitário de Santiago, 3810-193, Aveiro, Portugal}%

\author{Armando N. Pinto}
\affiliation{ Instituto de Telecomunicações, University of Aveiro, Campus Universitário de Santiago, 3810-193, Aveiro, Portugal}%
\affiliation{Department of Electronics, Telecommunications and Informatics, University of Aveiro, Campus Universitário de Santiago, 3810-193, Aveiro, Portugal}%

\date{\today}

\begin{abstract}
We present a polarization diverse receiver architecture for continuous variables quantum key distribution.
Our receiver architecture forgoes the need for any polarization matching calibrations or feedback loops and is usable in any locally generated local oscillator key distribution system.
We demonstrate experimentally that our system is capable of reliably achieving theoretical security against coherent attacks, even under very adverse random polarization drift scenarios, where the single polarization channels are unable to achieve security, and is capable of functioning for long periods of time.
\end{abstract}

\maketitle

\section{Introduction}
Continuous Variables Quantum Key Distribution (CV-QKD) tackles the problem of the generation and distribution of symmetric cryptographic keys without assuming any computational limitations while employing standard telecom equipment~\cite{grosshans02b}.
However, the amount of information available to an eavesdropper is highly dependent on the excess noise observed in the channel, which demands a careful and precise estimation of noise sources~\cite{laudenbach18}.
When implemented over standard optical fibres, one such noise source is random polarization drift in the communication channel, which will degrade the efficiency of the coherent detection scheme~\cite{liu20}.
Therefore a polarization drift compensation scheme is strictly necessary for the implementation of efficient and secure CV-QKD systems~\cite{wang19,zhao18,laudenbach18,pereira21}.
\par
Coherent-state CV-QKD typically encodes the information in the phase and amplitude of weak coherent states, thus allowing for implementation with current modulation methods and telecom-based equipment~\cite{grosshans02b,almeida21}.
The first implementations of CV-QKD protocols were carried out by using a transmitted local oscillator (LO) setup~\cite{ralph99}.
Nevertheless, that was found to be a security loophole, because an eavesdropper could manipulate the LO, thus hiding their tampering on the quantum signal itself~\cite{kleis17,laudenbach19}.
In that scenario, local LO (LLO) techniques, usually employing a relatively high power pilot tone aided by digital signal processing (DSP), are today the most common implementations of CV-QKD systems~\cite{kleis17,laudenbach19}.
Lately, LLO CV-QKD implementations using single-sideband modulation with true heterodyne detection have been proposed, avoiding low-frequency noise ~\cite{kleis17,laudenbach19}.
In order to further maximize noise rejection, CV-QKD implementations using root-raised-cosine (RRC) signal modulation have been explored~\cite{kleis17}.
Nevertheless, those implementations do not consider the impact of polarization mismatch between the quantum signal and the LLO.
%
Random polarization drift occurs naturally in fibres subjected to vibrations, temperature fluctuations, among others~\cite{liu19}.
Misalignments between the polarizations of the two laser fields interfering in the coherent detection scheme will severely reduce the efficiency of the detection scheme employed~\cite{liu20,kleis17}.
In CV-QKD communication systems, polarization drift is typically avoided, during a limited time window, by manually aligning the polarization of the signal with that of the LO~\cite{kleis17,laudenbach19}.
This may be appropriate in a laboratory environment, where stability times are typically in the range of hours~\cite{liu20}, but in field deployed fibres, especially aerially deployed ones, this stability will be on the order of minutes~\cite{liu19}.
Conversely, in classical communications, random polarization drift is compensated for by detecting both polarizations of the incoming light field and then compensating for the time-evolving drift in DSP~\cite{zhang12, faruk17}.
A system employing DSP aided polarization mismatch recovery was presented in~\cite{wang19}, using two optical hybrids coupled with four balanced coherent receivers.
%
\par
In this work, we present a polarization diverse receiver setup employing true heterodyne detection, requiring only two balanced coherent receivers, for use in CV-QKD applications, the first demonstration of such a scheme, to the best of our knowledge.
This experimental setup, coupled with the corresponding DSP, allows for passive polarization drift compensation, i.e. not requiring any manual tuning or feedback loop system.
We present experimental results showing that our system is able to achieve secure transmissions even in very adverse random polarization drift scenarios.
%
\begin{figure*}[t]
\centering
\includegraphics[width=\linewidth]{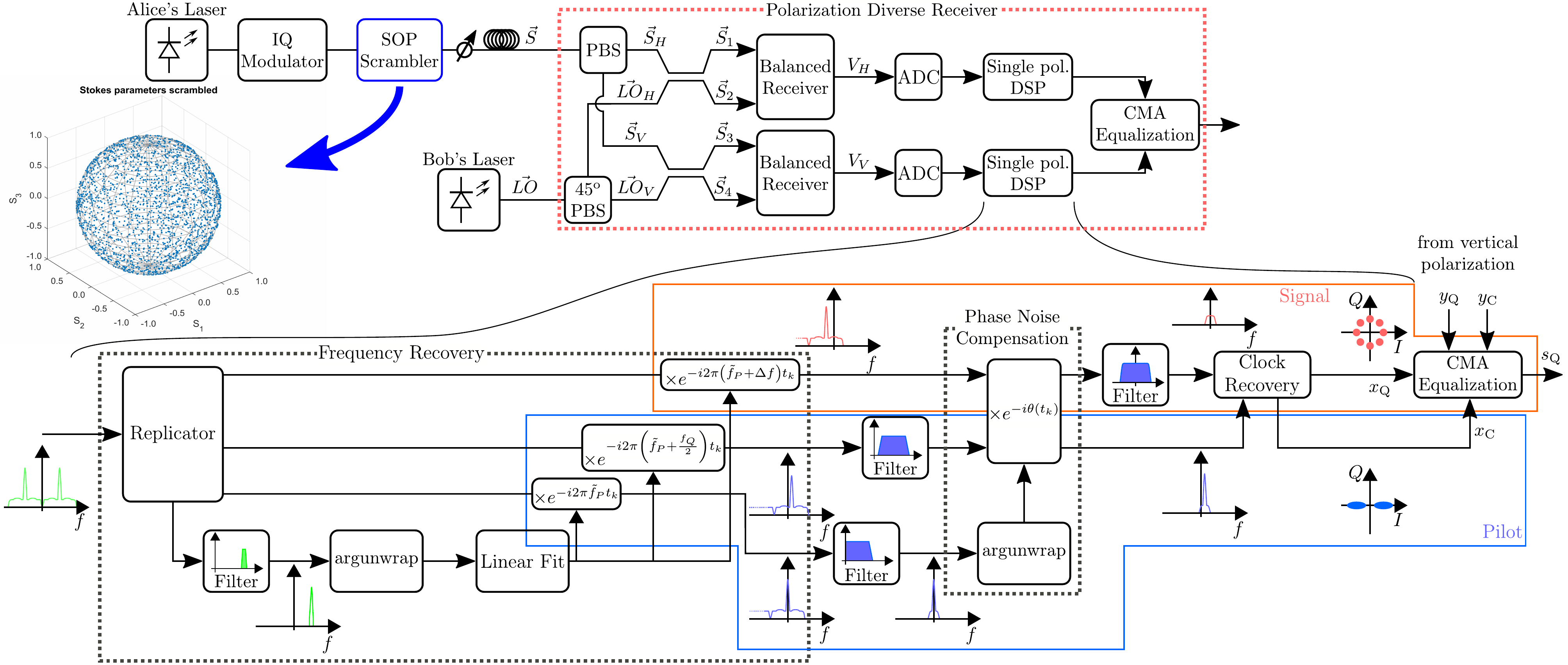}
\caption{Block diagram of the experimental system, the polarization diverse receiver system is highlighted.}
\label{fig:blockDiagram}
\end{figure*}
\par
This work is organized as follows.
We begin by fully describing our experimental system and the corresponding digital signal processing (DSP) utilized.
Secondly, we present and discuss experimental results extracted from the previously described system, showing the evolution of its estimated channel transmission, excess noise and secure key-rate.
We finalize this work with a summary of the major conclusions.

\section{System Description}
A block diagram of our system is presented in Figure~\ref{fig:blockDiagram}.
Alice starts by modulating the optical signal that she extracts from her local coherent source, which consists of a Yenista OSICS Band C/AG TLS laser, tuned to 1550.006~nm.
RRC modulation is chosen because of the possibility of using matched filtering at the receiver without inter-symbolic interference~\cite{faruk17}, thus allowing for optimum Gaussian white-noise minimization.
The symbol rate was set at 38.4~MBd, with an 8-phase-shift keying (8-PSK) constellation, the security of which, in the asymptotic regime, was established in~\cite{becir12}.
In order to avoid the high levels of noise present in the low frequency part of the electromagnetic spectrum~\cite{tomer60}, the RRC signal is up-converted in the transmitter to an intermediate frequency, $f_Q = 38.4$~MHz.
Furthermore, this signal is frequency multiplexed with a DC pilot tone, i.e. $f_P = 0$~Hz, which will be used for frequency and phase recovery at the receiver. 
This signal is fed into a Texas Instruments DAC39J84EVM digital to analog converter (DAC), which in turn drives a u2t Photonics 32 GHz IQ modulator coupled with a SHF807 RF amplifier.
The modulated signal is first passed through a Thorlabs PL100S State Of Polarization (SOP) Locker/Scrambler, which allows us to scramble the polarization state of the signal, and  then attenuated using a  Thorlabs EVOA1550F variable optical attenuator until the signal has on average 0.33 photons per symbol.
The signal is then sent through a single-mode fibre spool with length 40~km before arriving at the receiver.
At the receiver side, the signal is first passed through a PBS, splitting its polarizations and sending each to different 50/50 beam-splitters, where they are mixed with the LLO.
The LLO, which consists of a Yenista OSICS Band C/AG TLS laser tuned to 1549.999~nm.
In this situation the signals have a frequency shift of $f_S\approx1$~GHz, a value chosen to coincide with the flattest region of the balanced detectors' frequency response.
The LLO is also passed through a PBS, this one with its fast-axis shifted 45$^\text{o}$ in relation to the polarization alignment of the laser, effectively sending half the power to each individual 50/50 beam-splitter.
Both 50/50 beam-splitters are polarization maintaining, ensuring that the polarization of both the signal and LO mixed in each match.
The outputs of each 50/50 beam-splitter are fed into a pair of Thorlabs PDB480C-AC balanced optical receivers, connected to the inputs of a Texas Instruments ADC32RF45EVM ADC board, which is running at a sample rate of 2.4576 GS/s.
The digitized signal is then fed into the DSP stage, which is also presented in Figure~\ref{fig:blockDiagram}.
The bulk of the DSP is performed independently for each polarization, before the recovered constellations from each polarization are combined in a constant modulus algorithm (CMA) step.
\par
The DSP starts by performing frequency recovery, where four copies of the signal obtained from the ADC are taken and a tight digital pass-band filter, centered at $\tilde{f}_P = f_P + f_S$, is applied to one of them.
Extracting the phase from this filtered signal and fitting it against a time-vector will yield an estimation for $\tilde{f}_P$.
One of the other copies from the original signal is then downconverted by multiplying it by the complex oscillator $e^{-i2\pi\tilde{f}_Pt_k}$, where $t_k$ is a time-vector, thus placing the pilot signal at close to base band.
This signal will later be used for phase noise compensation.
The third copy of the original signal is downconverted  by another complex oscillator of the form $e^{-i2\pi\left(\tilde{f}_P+\frac{f_Q}{2}\right)t_k}$, which will cause the pilot to be located at roughly $\frac{f_Q}{2}$.
This signal will later be used for clock recovery.
The fourth and final copy of the original signal is downconverted by a third complex oscillator of the form $e^{-i2\pi\left(\tilde{f}_P+\Delta f\right)t_k}$, where $\Delta f=f_Q-f_P$, resulting in the oscillator taking the explicit form $e^{-i2\pi\left(f_Q+f_S\right)t_k}$, this places the quantum signal at close to base band.
Note that the estimation of $\tilde{f}_P$ is assumed to contain errors.
\par
The frequency compensated pilot and clock signals are then passed through a low-pass and a band-pass filter, respectively.
This filtering step will both reduce the noise present in the signals and isolate them from each other.
The phase of the filtered pilot signal, which is equal to the phase mismatch between the two lasers apart from a constant value, which in turn is obtained during an initial calibration stage, is then extracted and used to compensate for the phase noise in both the quantum signal and the clock. 
Since the pilot and signal are sampled at the same instant, the phase mismatch estimated from the former will equal that of the latter, thus residual phase noise will arise mainly from amplitude noise degrading the accuracy of the estimation~\cite{kleis17}.
The phase compensated quantum signal is then passed through its own matched filter. 
The filtering stage on the quantum signal is postponed until after the phase compensation step, this is done because small errors in the frequency estimate can be corrected by the phase noise compensation and application of the matched filter on the signal while it is not at base band may cause distortion in the final obtained constellation. 
\par
Finally, the filtered clock is used to re-sample both itself and the filtered quantum signal to one sample per symbol, with one sample being taken of each for every 0 of the imaginary component of the clock signal.
At the end of this clock recovery step we are in the possession of four constellations, two corresponding to the clock constellations of the clock signal, $x_\text{C}$ and $y_\text{C}$, and two to the quantum signal ones , $x_\text{Q}$ and $y_\text{Q}$.
\par
These four constellations are then fed into the CMA algorithm, which follows a slightly modified form of the method presented in~\cite{faruk17}.
Sliding blocks of N samples of each of the four constellations are isolated, taking the form of the column vectors
\begin{align}
\vec{x}_\text{Ci}(n) &= [x_\text{C}(n)~ x_\text{C}(n-1)~ ...~ x_\text{C}(n-N)]^T, \\
\vec{y}_\text{Ci}(n) &= [y_\text{C}(n)~ y_\text{C}(n-1)~ ...~ y_\text{C}(n-N)]^T, \\
\vec{x}_\text{Qi}(n) &= [x_\text{Q}(n)~ x_\text{Q}(n-1)~ ...~ x_\text{Q}(n-N)]^T, \\
\vec{y}_\text{Qi}(n) &= [y_\text{Q}(n)~ y_\text{Q}(n-1)~ ...~ y_\text{Q}(n-N)]^T. 
\end{align}
At the start of the algorithm, i.e. blocks $\vec{x}_\text{Ci,Qi}(0)/\vec{y}_\text{Ci,Qi}(0)$, are composed of all zeros except for the first element, which will consist of the first elements of the corresponding constellation.
The other elements of the sliding blocks are then progressively filled up.
The blocks for each signal are concatenated, resulting in the input column vectors~\cite{faruk17}
\begin{align}
\vec{u}_\text{Ci}(n) &= [\vec{x}_\text{Ci}(n);~\vec{y}_\text{Ci}(n)],\\
\vec{u}_\text{Qi}(n) &= [\vec{x}_\text{Qi}(n);~\vec{y}_\text{Qi}(n)].
\end{align}
Two N-tap filters are created, $\vec{h}_\text{x}$ and $\vec{h}_\text{y}$ , consisting also of column vectors.
At the start of the algorithm the first element of $\vec{h}_\text{x}$ and $\vec{h}_\text{y}$ is set to 1, with all the others being 0.
These two filters are concatenated,
\begin{align}
\vec{h} &= [\vec{h}_\text{x};~\vec{h}_\text{y}],
\end{align}
with the resulting filter being applied to the input column vectors following
\begin{align}
s_\text{C}(n) &= \vec{h}^\dagger\cdot \vec{u}_\text{Ci}(n),\label{eq:s_C}\\
s_\text{Q}(n) &= \vec{h}^\dagger\cdot \vec{u}_\text{Qi}(n),\label{eq:s_Q}
\end{align}
which correspond to the clock and quantum output constellations, respectively.
Note that both $\vec{h}$ and $\vec{u}_\text{Ci,Qi}(n)$ are $2N\times1$ column vectors, so for each of the inner products in~\eqref{eq:s_C} and~\eqref{eq:s_Q}, one output constellation point will be generated.
After each step $n$, the ``error", $\varepsilon$, of the algorithm is computed through~\cite{faruk17}
\begin{equation}
\varepsilon = \text{E}[|\vec{x}_\text{C}|]+\text{E}[|\vec{y}_\text{C}|] - s_\text{C}(n),
\end{equation}
which measures the distance of the amplitude of the latest output point of the clock constellation to the expected clock constellation amplitude.
This ``error" is then used to update the filter $h$ through~\cite{faruk17}
\begin{equation}
\vec{h} = \vec{h} + \mu \varepsilon s^*_\text{C}(n) \vec{u}_\text{C}(n)
\end{equation}
The output clock constellation can then be discarded, while the quantum output constellation is then be evaluated for security.
\par
\begin{figure}[h]
\centering\includegraphics[width=\linewidth, trim=4cm 8.5cm 4.7cm 8.6cm, clip]{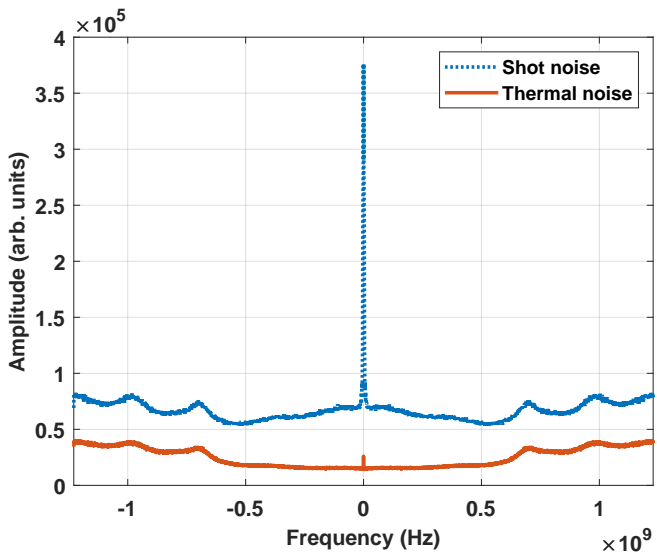}
\caption{Spectra of the thermal and shot noise snapshots taken from the experimental system.}
\label{fig:noiseSpectra}
\end{figure}

Protocol security is evaluated following the methodology presented in~\cite{becir12}.
The achievable secret key rate is given by
\begin{equation}\label{eq:keyRate}
K=\beta I_\text{BA}-\chi_\text{BE},
\end{equation}
where $\beta$ is the reconciliation efficiency, $I_\text{BA}$ is the mutual information between Bob and Alice, given by~\cite{becir12}
\begin{equation}
I_\text{BA} = \log_2\left(1+\frac{2\tilde{T}\eta\braket{n}}{2+\tilde{T}\eta\tilde{\epsilon}+2\epsilon_\text{thermal}}\right),
\end{equation}
where, in turn, $\tilde{T}$ is the estimate for the channel transmission, $\eta$ is the quantum efficiency of Bob's detection system, $\braket{n}$ is the average number of photons per symbol, $\tilde{\epsilon}$ is the estimate for the excess channel noise and $\epsilon_\text{thermal}$ is the receiver thermal noise, expressed in shot noise units (SNU).
In~\eqref{eq:keyRate}, $\chi_\text{BE}$ describes the Holevo bound that majors the amount of information that Eve can gain on Bob's recovered states, being obtained through equation (17) in~\cite{becir12}.
For the results presented in this work $\braket{n}$ was set at 0.33 photons per symbol, $\eta=0.72$ and $\tilde{\epsilon}$ and $\epsilon_\text{thermal}$ were dynamically estimated for each measurement run.
The shot and thermal noise estimations were made with recourse to a capture of the receiver output with the transmitter laser turned off and with both lasers turned off, respectively.
To obtain precise shot and thermal noise figures, the same DSP that was applied to the quantum signal was applied to the shot and thermal noise captures obtained previously, with the noise captures being down converted, phase compensated and filtered before their variance was computed.
This was necessary because both are highly dependent on their spectral position, as can be seen in their spectra, shown here in Figure~\ref{fig:noiseSpectra}.
Since we cannot measure the shot noise without also including the thermal noise, the latter was obtained first and its value was subtracted from the variance of the former, yielding an estimate for the true shot noise.
The variance of the shot and thermal noise signals, named here $\sigma^2_\text{shot}$ and $\sigma^2_\text{thermal}$ respectively, are both expressed in ADC counts.
Thermal noise is converted to SNU by dividing it by the shot noise estimate $\sigma^2_\text{shot}$, explicitly
\begin{equation}
\epsilon_\text{thermal}=\frac{\sigma^2_\text{thermal}}{\sigma^2_\text{shot}}.
\end{equation}
The signal output by Bob's DSP is also converted to SNU, this in turn is accomplished by dividing the ADC count output by $\sqrt{\sigma^2_\text{shot}}$.
Bob's and Alice's states, $b$ and $a$ respectively, are related by the normal linear model~\cite{kleis17}:
\begin{equation}
b = ta+z,
\end{equation}
where $a$ is assumed to be normalized such that ${\text{E}\lbrace|a|^2\rbrace = 1}$, $t = \sqrt{\eta T 2\braket{n}}$ and $z$ is the noise contribution, which follows a normal distribution with null mean and variance ${\sigma^2=2+2\epsilon_\text{thermal}+\eta T \epsilon}$.
$t$ and $\sigma^2$ can be estimated through~\cite{kleis17}
\begin{equation}
\tilde{t} = \text{Re}\left\lbrace\frac{\sum_{i=1}^Na_ib_i^*}{N}\right\rbrace,\qquad
\tilde{\sigma^2}=\frac{\sum_{i=1}^N|b_i-\tilde{t}a_i|^2}{N},
\end{equation}
the transmission and excess noise are then estimated through
\begin{equation}
\tilde{T} = \frac{t^2}{\eta2\braket{n}},\qquad\tilde{\epsilon}=\frac{\sigma^2-2-2\epsilon_\text{thermal}}{\eta \tilde{T}}.
\end{equation}
For a truly secure communication, the uncertainty of the channel parameter estimations needs to be taken into account, choosing the confidence bound that gives the most advantage to Eve~\cite{leverrier10}.
As the objective of this letter is to show the capabilities of our polarization diverse receiver, only the central estimate for the parameters is used. 

\section{Experimental Results}
\begin{figure}[h]
\centering\includegraphics[width=\linewidth, trim=3.9cm 8.5cm 4.6cm 9.1cm, clip]{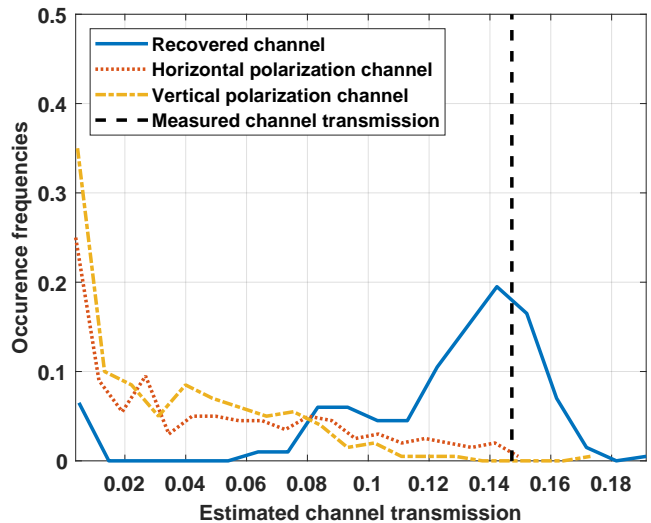}
\caption{Distribution of the estimated channel transmissions for both the single polarization and recovered channels in the scrambled scenario.}
\label{fig:channelTransmission}
\end{figure}

\begin{figure}[h]
\centering\includegraphics[width=\linewidth, trim=4.2cm 8.4cm 4.5cm 9cm, clip]{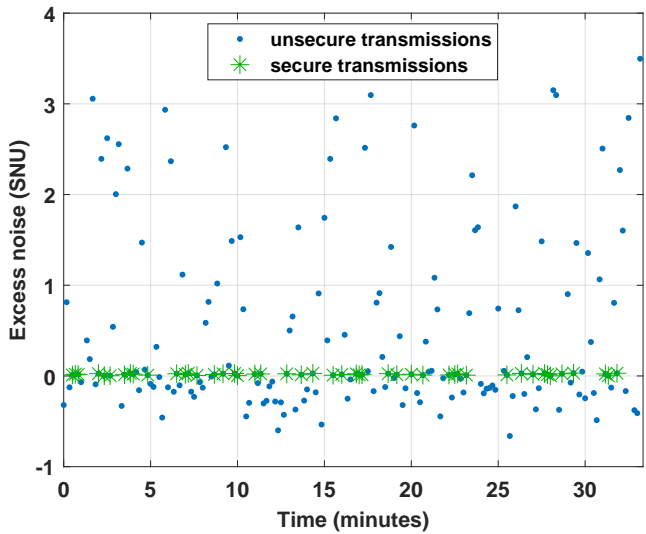}
\caption{Evolution of the estimated excess noise for the 200 results taken. Situations where the a secure key was able to be transmitted are highlighted with green asterisks. Results taken in the scrambled scenario.}
\label{fig:excessNoise}
\end{figure}

\begin{figure}[h]
\centering\includegraphics[width=\linewidth, trim=3.8cm 8.4cm 4.5cm 9cm, clip]{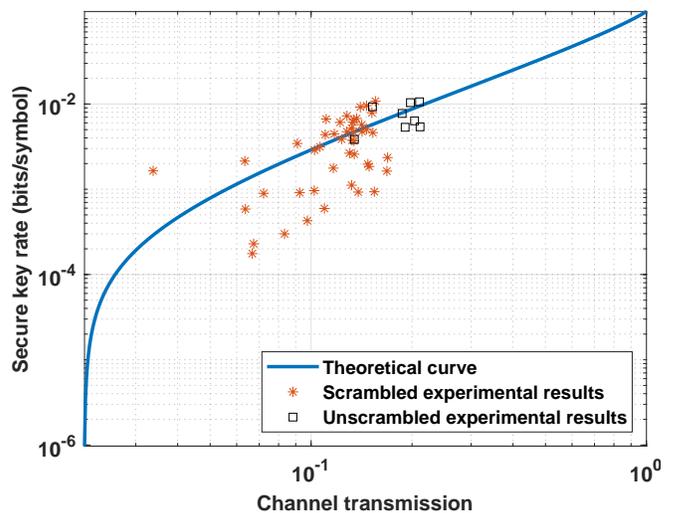}
\caption{Achievable key rate, given by~\eqref{eq:keyRate}, for our polarization diverse receiver, with $\beta=0.95$.}
\label{fig:secureKeyRate}
\end{figure}

The system was run freely for a half hour in scrambled mode and another half hour in unscrambled mode, with 2~ms snapshots taken every 10 seconds. 
A total of 200 snapshots were taken in each scenario, each snapshot containing 65536 symbols.
The occurrence frequencies for the channel transmission estimates in the scrambled scenario, for both the recovered and the single-polarization channels, is presented in Figure~\ref{fig:channelTransmission}, where a line indicating the empirically determined channel transmission is also included.
From the results in Figure~\ref{fig:channelTransmission} we can see that the single polarization channels exhibit a much lower transmission on average, with the most likely values being located nowhere near the actual value of the channel transmission.
Meanwhile, the distribution of the estimated channel transmission from the recovered channel exhibits a maximum very close to the transmission determined empirically, with the observed deviation being attributable to losses in the receiver and in the fibre connectors.
In Figure~\ref{fig:excessNoise} we present the values of the estimated excess noise in the recovered channel observed for each of the 200 results taken in the scrambled mode.
Situations where the a secure key was able to be transmitted are highlighted with green asterisks.
We can see that our system was able to recover secure keys for the duration of the experiment, with excess noise hovering quite close to 0, apart from some deviations caused by failures in signal recovery, coinciding with the cases with very low transmission in the recovered channel observed in Figure~\ref{fig:channelTransmission}.
We have all reasons to believe that the system could've been run for an indefinite amount of time and still be able to achieve secure transmissions.
Further optimization of the noise calibration step could improve the overall efficiency of the system.
Some situations exhibit negative excess noise, these can be attributed to fluctuations of the thermal and shot noises, causing the variance observed during the snapshot to be lower than at the noise estimation steps.
Time-evolving imbalances of the optical components could also be a contributing factor~\cite{pereira21}.
Finally, we show the experimentally observed secure key rates in function of channel transmission in Figure~\ref{fig:secureKeyRate}, alongside with the corresponding theoretical curve, for which the average values of the observed excess noise and thermal noise were used.
Data from both scrambled and unscrambled polarization scenarios is included.
We can see that our experimental results closely adhere to the theoretical curve.
The small separation between the scrambled and unscrambled results can be attributed to the results being taken in different days, thus having slightly different conditions (for example in temperature).
We see that in both scenarios we were able to achieve secure key rates of roughly 0.01~bits/symbol.
No secure transmissions were observed for the individual polarization channels.

\section{Conclusion}
In summary, we present a polarization diverse receiver architecture that avoids the need for manual calibration or complex feedback loops to recover from random polarization drift.
Our system works by passively monitoring both polarizations continuously and recovering the full channel from the single polarization ones.
Our system was capable of working for an indefinite period of time at a transmission distance compatible with metro network connections.
Furthermore, this stability is achieved with a relatively simple and inexpensive receiver design.
We believe our contribution brings CV-QKD closer to widespread adoption.

\begin{acknowledgments}
This work was supported in part by Fundação para a Ciência e a Tecnologia (FCT) through national funds, by the European Regional Development Fund (FEDER), through the Competitiveness and Internationalization Operational Programme (COMPETE 2020) of the Portugal 2020 framework, under the PhD Grant SFRH/BD/139867/2018, projects Q.DOT (POCI-01-0247-FEDER-039728), UIDB/50008/2020-UIDP/50008/2020 (action QuRUNNER and QUESTS).
\end{acknowledgments}

\bibliography{bibliography}

\end{document}